\documentclass[twocolumn,showpacs,preprintnumbers,amsmath,amssymb]{revtex4}
\usepackage{graphicx}
\usepackage{dcolumn}
\usepackage{bm}

\begin{document}

\preprint{APS/123-QED}
\title{Observation of cooperative Mie scattering from an ultracold atomic cloud}

\author{H. Bender$^a$}
\author{C. Stehle$^a$}
\author{S. Slama$^a$}
\author{R. Kaiser$^c$}
\author{N. Piovella$^b$}
\author{C. Zimmermann$^a$}
\author{Ph.W. Courteille$^{acd}$}
\affiliation{$^a$Physikalisches Institut, Eberhardt-Karls-Universit\"at T\"ubingen, D-72076 T\"ubingen, Germany.\\
	$^b$Dipartimento di Fisica, Universit\`a Degli Studi di Milano, Via Celoria 16, I-20133 Milano, Italy.\\
	$^c$Institut Non Lin\'eaire de Nice, CNRS, Universit\'e de Nice Sophia-Antipolis, F-06560 Valbonne, France.\\
	$^d$Instituto de F\'isica de S\~ao Carlos, Universidade de S\~ao Paulo, 13560-970 S\~ao Carlos, SP, Brazil.}

\date{\today}

\begin{abstract}
Scattering of light at a distribution of scatterers is an intrinsically cooperative process, which means that the scattering rate and the angular distribution of the scattered light are essentially governed by bulk properties of the distribution, such as its size, shape, and density, although local disorder and density fluctuations may have an important impact on the cooperativity. Via measurements of the radiation pressure exerted by a far-detuned laser beam on a very small and dense cloud of ultracold atoms, we are able to identify the respective roles of superradiant acceleration of the scattering rate and of Mie scattering in the cooperative process. They lead respectively to a suppression or an enhancement of the radiation pressure. We observe a maximum in the radiation pressure as a function of the induced phase shift, marking the borderline of the validity of the Rayleigh-Debye-Gans approximation from a regime, where Mie scattering is more complex. Our observations thus help to clarify the intricate relationship between Rayleigh scattering of light at a coarse-grained ensemble of individual scatterers and Mie scattering at the bulk density distribution.
\end{abstract}

\maketitle

At first sight Rayleigh scattering from point-like particles and Mie scattering from extended objects are quite different phenomena. For example, Rayleigh scattering exhibits resonances due to the internal energy structure of the particles, e.g.~atoms, while Mie scattering shows resonances induced by the boundary conditions, the target imposes to the incident light field \cite{vandeHulst-81,Bohren-83}. On the other hand, any extended object, e.g.~a dielectric sphere, is an assembly of microscopic scattering particles. This raises questions concerning the description of Mie scattering at objects with coarse-grained density distributions. Another interesting question is, whether small ensembles of particles exhibit scattering features normally attributed to Mie scattering.

While Mie scattering is obviously a cooperative process involving a macroscopic particle distribution, in quantum optics Rayleigh scattering is often described as a single particle effect. That this concept is erroneous in the case of ensembles of scatterers, has been recognized as early as 1954 by Dicke \cite{Dicke54}. Rather, collective effects play a dominant role even on the level of single-photon scattering \cite{Scully06,Eberly06,Svidzinsky08,Courteille10} provided, as has been demonstrated in a recent experiment \cite{Bienaime10}, the coarse graining of the cloud's density distribution is negligible.

The scattering of light is usually studied via its angular distribution. However if the target is an atomic cloud with temperature below the recoil limit, the modification of the atomic momentum distribution due to radiation pressure can conveniently be mapped by time-of-flight imaging techniques and may yield valuable information about the nature of the scattering process. Momentum images are routinely analyzed in experiments with ultracold or Bose-condensed atoms, however mostly in regimes where local atomic disorder dominates over global Mie scattering. In this letter, we present an experimental approach emphasizing the role of Mie scattering. The idea consists in using ultracold atomic clouds compressed to volumes so small, that the whole cloud acts itself as an inhomogeneity at which light is diffracted like at a very small dielectric sphere. Furthermore, instead of analyzing the whole momentum distribution, which is blurred by the frequent occurrence of interatomic collisions in the dense cloud, we concentrate on its first moment, i.e.~the center-of-mass acceleration under the influence of radiation pressure. We discover, as a clear signatures of Mie scattering, an up to three-fold increase of the cooperative radiation pressure with respect to single-atom radiation pressure. Exploiting the fact that the index of refraction of an atomic cloud can be tuned over huge ranges by changing the cloud's density and volume, or by tuning the light frequency, we are able to address the Rayleigh-Debye-Gans limit of small phase-shifts for the incident laser beam or to enter the Mie regime of large phase shifts.

\bigskip
\textit{Force on the center-of-mass.---} We start with a brief reminder of the model of collective radiation pressure exerted by a low intensity incident laser beam on a cloud of $N$ two-level atoms \cite{Courteille10,Bienaime10} concentrating on cases of large push beam detunings, $\Delta_0\gg\Gamma$, where multiple scattering and collective saturation can be neglected. In this limit, the modification of the radiation pressure $F_c$ acting on the center-of-mass of the ensemble of atoms with respect to the radiation pressure $F_1$ acting on isolated atoms can be expressed as \cite{Courteille10,Bienaime10},
\begin{equation}\label{EqRadPresN}
	\frac{F_c}{F_1}=\frac{4\Delta_0^2}{4\Delta_0^2+N^2\Gamma^2s_N^2}\times N\left(s_N-f_N\right)~.
\end{equation}
The averages $s_N=\langle|S_N|^2\rangle_{\theta,\varphi}$ and $f_N=\langle|S_N|^2\cos\theta\rangle_{\theta,\varphi}$ of the structure factor $S_N(\mathbf{k})=\frac{1}{N}\sum_{j=1}^Ne^{(\mathbf{k}-\mathbf{k}_0)\cdot\mathbf{r}_j}$ are taken over the total solid angle of emission of a photon with wavevector $\mathbf{k}$, at an angle $\theta$ with the incident wavevector $\mathbf{k_0}$ \cite{Courteille10}. They account for the role of the geometry in the photon absorption and reemission process, respectively. As stated in Ref.~\cite{Bienaime10}, they are well approximated by $s_N\simeq1/N+s_\infty$ and $f_N\simeq f_\infty$, where $s_\infty$ and $f_\infty$ are calculated for a smooth density distribution, i.e.~neglecting disorder and coarse-graining in the atomic cloud. The averages $s_\infty$ and $f_\infty$ have been calculated in Ref.~\cite{Courteille10} for ellipsoidal Gaussian density distributions, $n(\mathbf{r})=n_0\exp[-(x^2+y^2)/2\sigma_r^2-z^2/2\sigma_z^2]$, with the size $\sigma_r=\sigma/k$ and the aspect ratio $\eta=\sigma_z/\sigma_r$.

\bigskip
\textit{Further approximations.---} In order to simplify the interpretation of Eq.~(\ref{EqRadPresN}), we restrict ourselves to regimes of negligible multiple scattering, $4\Delta_0^2/\Gamma^2>b_0$, where $b_0\equiv3N/\sigma^2$ is the resonant optical density. This condition implies the inequality 
\begin{equation}\label{EqIneq}
	4\Delta_0^2/\Gamma^2>Ns_\infty~,
\end{equation}
since for large ($\sigma\gg 1$) spherical clouds $s_\infty\equiv s_\infty^{(\eta=1)}=(2\sigma)^{-2}$ and for elongated clouds $s_\infty^{(\eta>1)}<s_\infty^{(\eta=1)}$ \cite{Courteille10}. Using the inequality (\ref{EqIneq}), we can approximate the radiation pressure modification (\ref{EqRadPresN}) as
\begin{equation}\label{EqRadPres1}
	\frac{F_c}{F_1}\simeq\frac{4\Delta_0^2}{4\Delta_0^2+N^2\Gamma^2s_\infty^2}\times \left(1+Ns_\infty-Nf_\infty\right)~.
\end{equation}
The fraction in this equation describes the role of the superradiant enhancement of the scattering rate $\Gamma\longrightarrow N\Gamma s_N$. The fact that the fraction is obtained from the fraction in Eq.~(\ref{EqRadPresN}) by substituting the factor $s_N$ by $s_\infty$ signifies that, in the single scattering regime and for large detunings, $\Delta_0\gg\Gamma$, disorder does not influence the superradiant acceleration of scattering significantly. In contrast disorder does shape the scattering through the bracket in Eq.~(\ref{EqRadPres1}), which represents a geometrical factor accounting for the angular distribution of the \textit{reemitted} light. This bracket is responsible for Mie scattering, as we will see in the following.

\begin{figure}[h]
		\includegraphics[width=8.7cm]{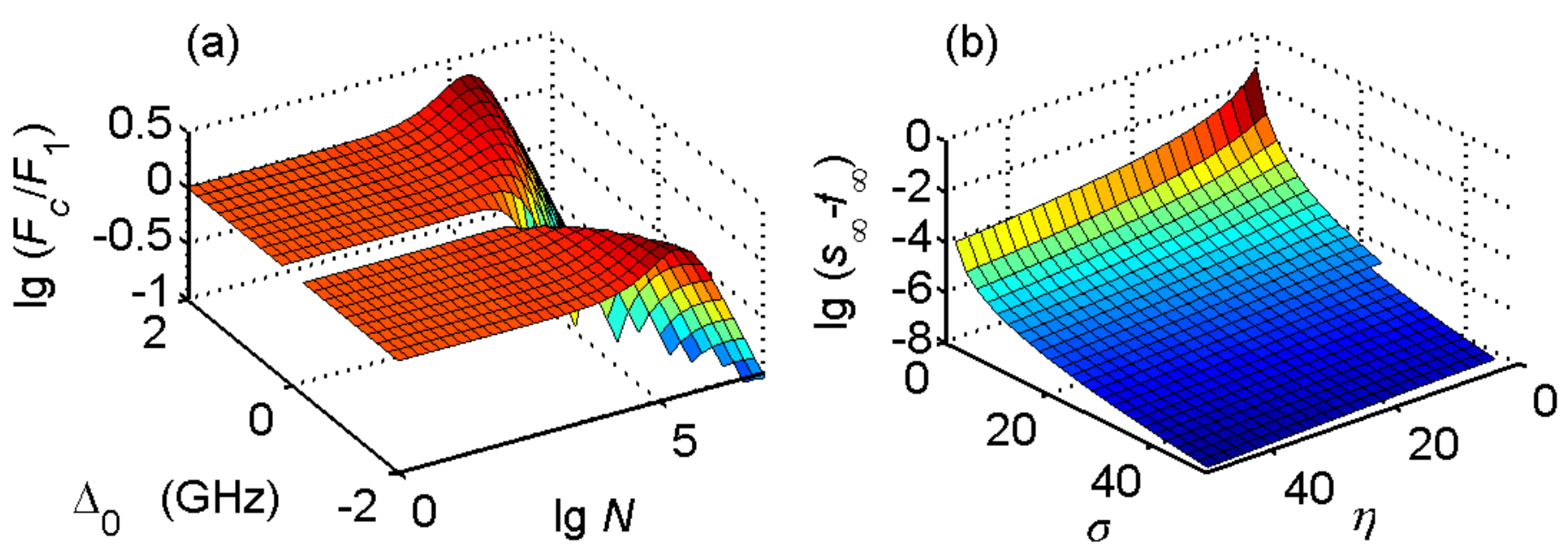}
		\caption{(color online) (a) Radiation pressure modification as a function of $\Delta_0$ and $N$. Here $\sigma=10.4$ and $\eta=6.3$.
			(b) Dependence of $s_\infty-f_\infty$ on $\sigma$ and $\eta$.}
		\label{FigForce}
	\end{figure}

Fig.~\ref{FigForce}(a) shows the dependence of the radiation pressure modification on $\Delta_0$ and $N$. The multiple scattering regime to be avoided lies in the narrow gap separating the regimes of red and blue detuning.

\bigskip
\textit{Reduction and enhancement of radiation pressure.---} For large cloud volumes, $\sigma\gg1$, the inhomogeneity represented by the shape of the cloud is negligible, so that reemission occurs predominantly into forward direction. Similarly, elongated clouds, $\eta\gg1$, bundle the emitted light into forward direction. Both features result in $s_\infty-f_\infty\simeq0$. Hence, the geometrical factor is dominated by disorder, and the radiation pressure modification is well approximated by just the fraction of Eq.~(\ref{EqRadPres1}). In this regime, the role of collectivity is limited to a superradiant enhancement of the decay rate $\Gamma$. In that case, the radiation pressure $F_c/F_1$ drops monotonically with increasing $N$. Hence, for extended clouds there is only \emph{cooperative reduction} of the radiation pressure with respect to uncorrelated radiation pressure. Measurements of this effect have already been published in Refs.~\cite{Bienaime10}.

\bigskip
In order to get noticeable \emph{cooperative radiation pressure enhancement}, two conditions must be fulfilled. Firstly, the fraction in Eq.~(\ref{EqRadPres1}) has to be close to unity, $2\Delta_0\gg N\Gamma s_\infty$, and secondly the bracket must be larger than unity, $N(s_\infty-f_\infty)>1$. The first condition is true for sufficiently large detunings or small atom numbers. In this case, the radiation pressure modification is well approximated by just the bracket of Eq.~(\ref{EqRadPres1}). The dependency of $s_\infty-f_\infty$ on $\sigma$ and $\eta$ is plotted in Fig.~\ref{FigForce}(b). The figure shows that smaller cloud sizes yield larger $s_\infty-f_\infty$. Furthermore, for a strong radiation pressure enhancement, spherical clouds are more advantageous than elongated ones. The bracket of Eq.~(\ref{EqRadPres1}) describes the role of cooperative Mie scattering and its interplay with disorder. In the regime where the bracket is considerably greater than 1, Mie scattering overwhelms the impact of disorder. The observation of an increase of the radiation pressure above the single-atom value thus constitutes a clear indication of Mie scattering, which has never been observed before.

The relative importance of the two effects, superradiance or Mie scattering, can be adjusted via judicious choice of the push beam detuning, $\Delta_0$, with respect to $N\Gamma s_\infty$. This allows us to address both limits in the following measurements.

\bigskip
\textit{Experiment.---} In principle, reducing the volume of a trapped gas to sizes not much larger than an optical wavelength is just a matter of cooling and compression. However, the compression of atomic clouds to very small volumes is not trivial, because the increased density leads to high three-body collision rates. Furthermore, below a critical temperature, the cloud crosses the threshold to Bose-Einstein condensation and develops a repulsive mean-field, which resists compression \cite{NoteCompCloud}.
	\begin{figure}[h]
		\includegraphics[width=6cm]{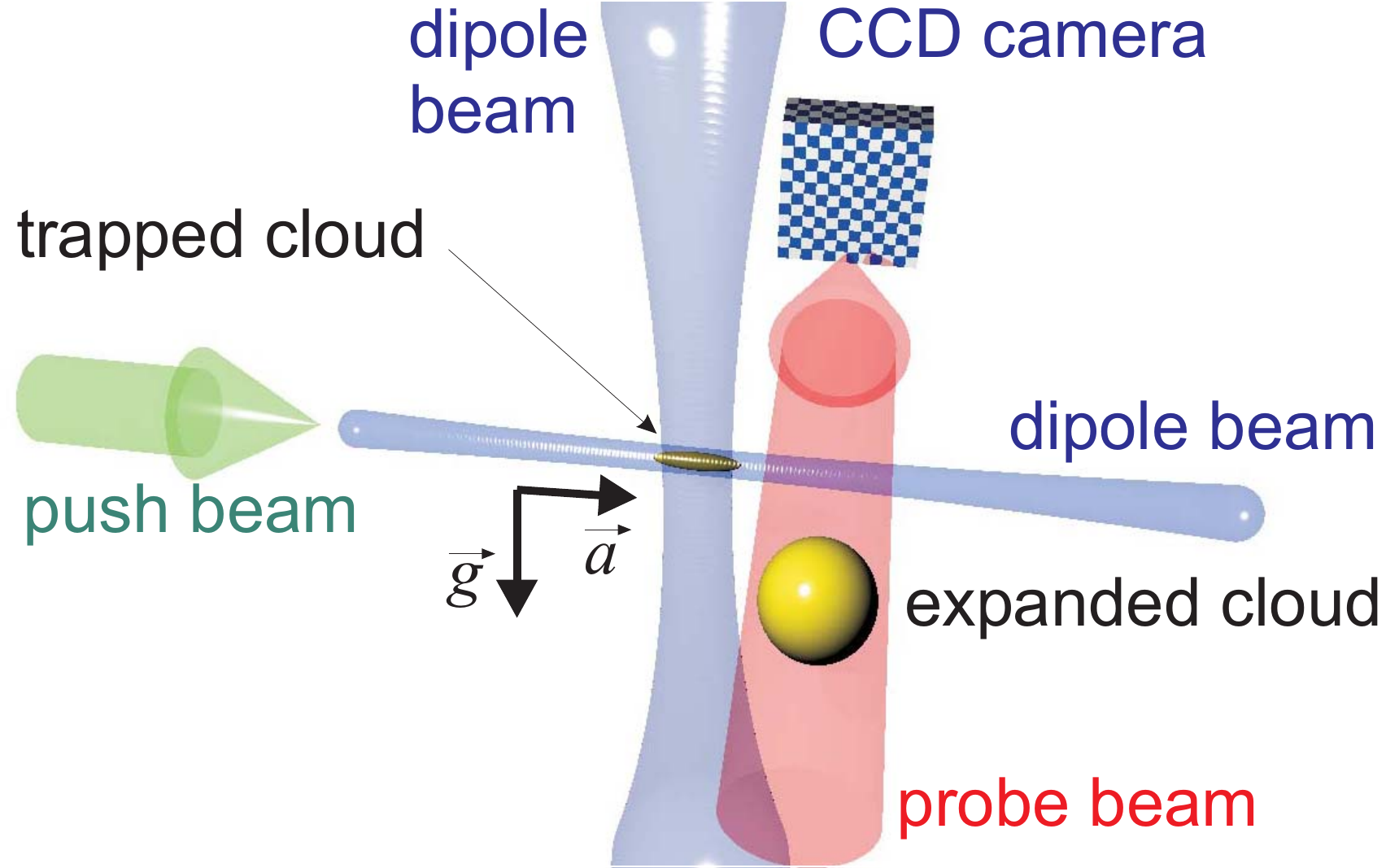}
		\caption{(color online) Scheme of the experiment showing the dipolar trapping beams, the push beam, and the imaging beam. The directions of radiation
			pressure, $\vec a$, and gravity, $\vec g$, are also shown.}
		\label{FigExpSetup}
	\end{figure}

We prepare a $^{87}$Rb atomic cloud in the $|F,m_F\rangle=|2,2\rangle$ hyperfine state in a magnetic trap and evaporatively cool it down to quantum degeneracy. Typically we have $N=10^4..10^6$ atoms at temperatures between $T=100..1000~$nK. The atom number is controlled by manipulating the initial atom number before applying the evaporation ramp. The cold atomic cloud is now loaded into a crossed beam dipole trap by ramping up the light fields within 100~ms, and then slowly ramping down the magnetic trap. The horizontal beam of the crossed dipole trap is generated by a fiber laser at $\lambda_r=1080~$nm. The power of the horizontal beam is $P_r=2~$W and its waist $w_r=25~\mu$m. The corresponding trap depth is $U_r=\frac{3\pi c^2}{2\omega_0^3}\frac{\Gamma}{\Delta_r}I_h \simeq k_B\times250~\mu$K and the trap frequency is $\omega_r=4U_r/mw_r^2=(2\pi)~1900~$Hz. The vertical beam is generated by a titanium-sapphire laser at $\lambda_z=820~$nm. Its power is $P_z=230~$mW and its waist $w_z=100~\mu$m. The corresponding trap depth is $U_z\simeq k_B\times90~\mu$K and the trap frequency $\omega_z=(2\pi)~300~$Hz. The aspect ratio of the trap is thus $\eta=6.3$.

The temperature of the cloud after evaporation depends on the atom number. However upon transfer into the dipole trap, the cloud is non-adiabatically heated to a temperature of $T\simeq2.5~\mu$K, which is almost independent from the atom number. For that reason the cloud is thermal. From this, we estimate a cloud radius of $\sigma=k\sqrt{k_BT/m\omega_r^2}\simeq10.4$.

To perform the measurement, we now apply along the weak trapping (axial) direction $\omega_z$ a $\sigma_+$-polarized light pulse of duration $\tau_0=20~\mu$s and intensity $I_0=95..730~\text{mW/cm}^2$ detuned from the D2 line by $\Delta_0/2\pi=0.5..4~$GHz. Immediately after the pulse (within $100~\mu$s) the trap is switched off. The cloud falls in free expansion for $\tau_{tof}=20~$ms before we apply an imaging pulse in transversal (radial) direction.
	\begin{figure}[h]
		\includegraphics[width=8.5cm]{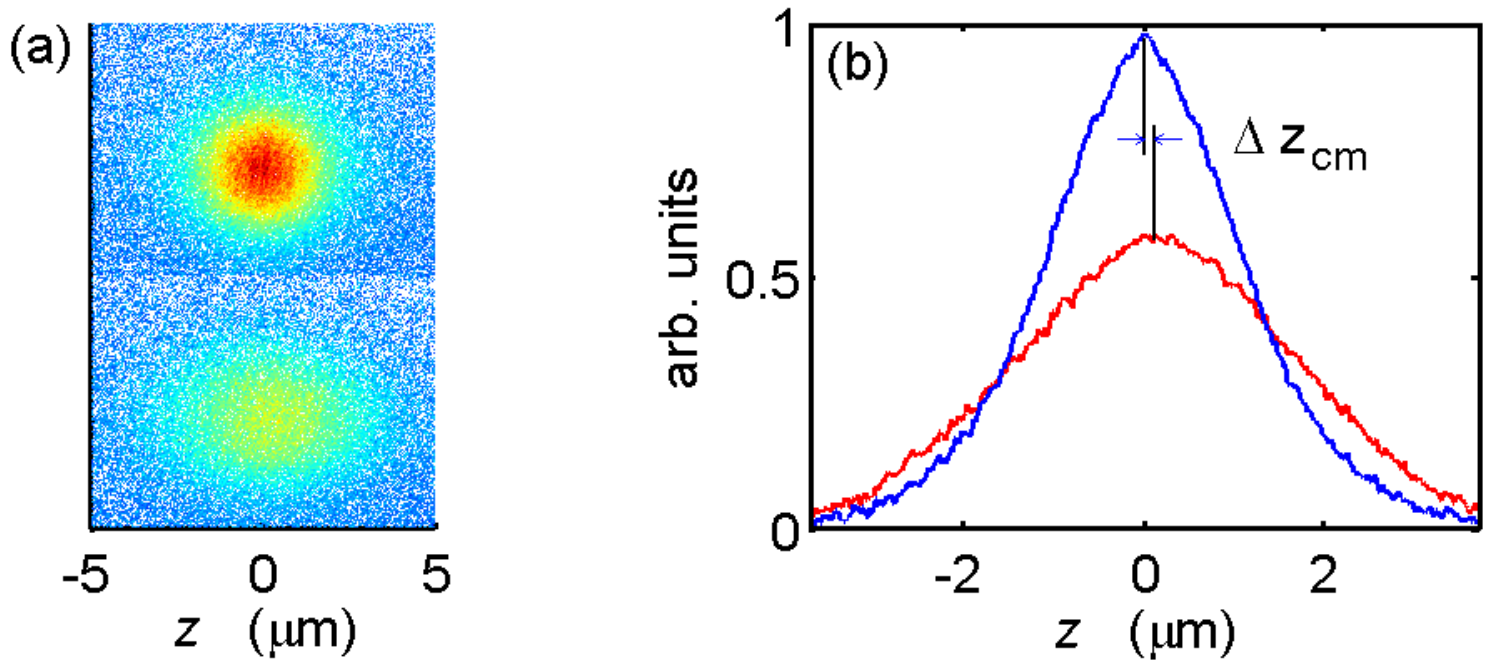}
		\caption{(a) Typical absorption images taken on a cloud released from the dipole trap without (upper half) and with (lower half) push beam
			sequence. Note that about 16\% of the atoms are lost under the action of the push beam.
			(b) Vertical integration of the images shown in (a). The flatter curve, which corresponds to the cloud after the push beam sequence, is shifted
			towards the right by an amount $\Delta z_{cm}$.}
		\label{FigFoto}
	\end{figure}

\bigskip
\textit{Observations.---} Fig.~\ref{FigFoto}(a) shows a typical time-of-flight absorption image reflecting the momentum distribution of the trapped atomic cloud. With $N=5\cdot10^5$ atoms we expect high densities of $n_0\simeq4\cdot10^{15}~\text{cm}^{-3}$ and large collision rates of $\gamma_{coll}\simeq25000~$s$^{-1}$. I.e.~during the time $\tau_0$ every atom undergoes almost one collision, which smears out and broadens the momentum distribution. This does however not affect the center-of-mass momentum imparted to the atomic cloud.

We compare the center-of-mass positions of absorption images of the atomic cloud with and without push beam sequence and determine their relative displacement $\Delta z_{cm}$ [see Fig.~\ref{FigFoto}(b)]. We estimate the collective radiation pressure from
\begin{equation}
	F_c=\frac{m\Delta z_{cm}}{\tau_0\tau_{tof}}
\end{equation}
and the single-atom radiation pressure from
\begin{equation}
	F_1=\hbar k \frac{I_0}{\hbar\omega} \frac{3\lambda^2}{2\pi}\frac{\Gamma^2}{4\Delta^2+\Gamma^2}~.
\end{equation}

\begin{figure}[h]
		\includegraphics[width=8.5cm]{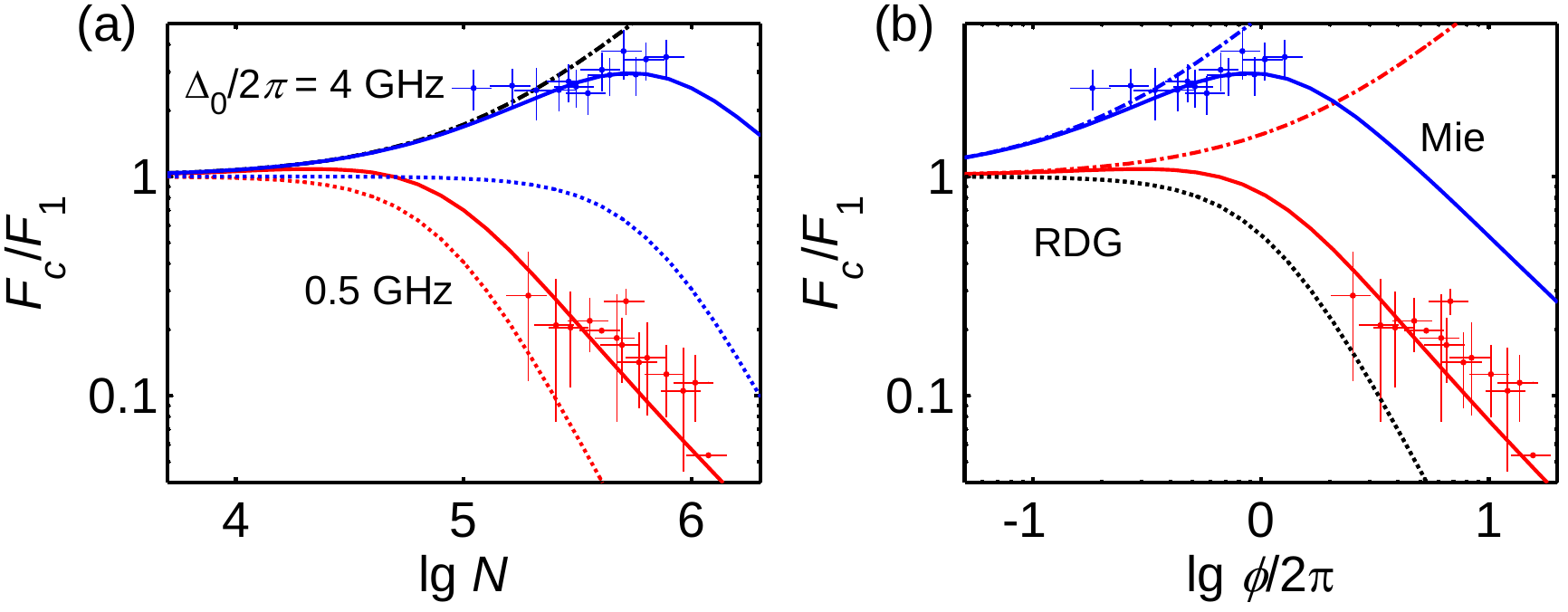}
		\caption{(color online) (a) Double-logarithmic plot of the measured (symbols) and calculated (lines) $N$-dependence of radiation pressure.
			The red symbols and lines correspond to the detuning $\Delta_0=(2\pi)~500~$MHz and the intensity $I_0=95~\text{mW/cm}^2$.
			For the blue symbols and lines, $\Delta_0=(2\pi)~4~$GHz and $I_0=730~\text{mW/cm}^2$.
			The red and blue solid lines show calculations based on the full radiation pressure expression~(\ref{EqRadPres1}) without adjustable parameters.
			The red and blue dotted lines represent just the fraction of the expression. The dash-dotted lines representing just the bracket coincide for 
			both values of $\Delta_0$.
			The horizontal error bars result from uncertainties in the measurement of the atom number and also from the fact that, during the push time,
			$\tau_0$ the atoms are pumped into a dark hyperfine state. The vertical error bars include an assumed uncertainty of 20\% for the push beam
			intensity $I_0$.
			(b) Same radiation pressure data as (a), but plotted as a function of the push beam phase shift $\phi$. In this diagram the
			dotted lines representing the fraction coincide for both values of $\Delta_0$.}
		\label{FigDipolfalle}
	\end{figure}

The calculated $N$-dependence of the cooperative modification of the radiation pressure force acting on the center-of-mass of a trapped cloud, Eq.(\ref{EqRadPres1}), is compared to measurement in Fig.~\ref{FigDipolfalle}(a). The $N$-dependence is flat in the disorder-dominated regime of low atom numbers. For larger atom numbers, the force increases first, exhibits a maximum and drops for large atom numbers. For increased detunings $\Delta_0$, the maximum shifts toward larger atom numbers and becomes more pronounced. For $\Delta_0=(2\pi)~0.5~$GHz the experimentally accessible atom numbers, $N=10^4..10^6$, are in a regime, where radiation pressure is strongly reduced by up to a factor of 10 [see red solid curve and symbols in Fig.~\ref{FigDipolfalle}(a)]. For $\Delta_0=(2\pi)~4~$GHz we expect an increased radiation pressure, i.e.~$F_c/F_1$ is larger than one, which is confirmed by the data [see blue solid curve and symbols in Fig.~\ref{FigDipolfalle}(a)]. 

During the push beam irradiation, a fraction of typically 20\% of the atoms is lost due to hyperfine pumping into the $F=1$ state. This is caused by a misalignment of the push beam with respect to the magnetic field direction resulting in a deviation of the polarization from perfect $\sigma_+$. The resulting uncertainty in atom number is illustrated by the horizontal error bars in Figs.~\ref{FigDipolfalle}. Also shown in Figs.~\ref{FigDipolfalle} are the curves representing the separate contributions of superradiance and Mie scattering. Obviously, the data corresponding to $\Delta_0=(2\pi)~0.5~$GHz are better explained by the impact of superradiance, while the data corresponding to $\Delta_0=(2\pi)~4~$GHz can be interpreted as being due to Mie scattering.

For the push beam the atomic cloud represents a dielectric medium whose mean refraction index can be approximated by $n_{rf}=1-\phi/\sigma+ib/2\sigma$, where the off-resonant optical density is $b=b_0\Gamma^2/(4\Delta^2+\Gamma^2)$ and the phase shift $\phi=b\Delta_0/\Gamma$.
Plotting the radiation pressure force as a function of the push beam phase shift [see Fig.~\ref{FigDipolfalle}(b)], we infer that the data taken at large detunings are in the so-called Rayleigh-Debye-Gans regime of Mie scattering, characterized by small phase shifts $\phi<2\pi$. The data taken at small detunings are in the Mie regime of large phase shifts, $\phi>2\pi$. However in this regime, the observed rapid decrease with $\phi$ of the radiation pressure is dominated by superradiant enhancement of scattering and not by Mie scattering. This is reflected in the absence of a maximum in the $\phi$-dependence of the radiation pressure. The appearance of a maximum for large detunings suggests a strong relationship to the first Mie resonance \cite{Sorensen00,Berg05}, which needs however to be confirmed by a more detailed theoretical model. The model, which is presently under investigation, would have to include the phase shift for the push beam induced by the cloud's refractive index. We expect that this model should also be able to predict the occurrence of high-order Mie resonances, which could be experimentally observed in even more compact atomic clouds.

\bigskip
\textit{Conclusion.---} We have presented measurements on ultracold atoms proving that, in the limit of the approximation made, the expression (\ref{EqRadPres1}) correctly describes the radiation pressure force on atomic clouds. More generally, it applies to extended objects that can either be ensembles of scatterers like homogeneous, ordered or disordered atomic clouds of arbitrary shapes and volumes, or macroscopic objects like dielectric spheres. This expression thus represents a bridge between microscopic Rayleigh scattering and macroscopic Mie scattering. Nevertheless, at high densities, near-field effects may come into play, calling for further corrections to the present simple model.

Atomic clouds with small atom numbers basically represents a randomly distributed bunch of scatterers, whose intrinsic disorder spoils cooperativity. The radiation pressure is then well described by the single-atom value. At large atom numbers, the atomic cloud forms a smooth density distribution characterized by an almost perfect cooperativity. In this regime, depending on the laser detuning, the cooperativity is either dominated by superradiant acceleration of the decay rate, for which case we experimentally observed an up to 10-fold reduction of the radiation pressure. Or it is dominated by Mie scattering leading to an observed up to 3-fold enhancement of the radiation pressure.

\bigskip
This work has been supported by the Deutsche Forschungsgemeinschaft (DFG) under Contract No. Co~229/3-1.


\begin{thebibliography}{10}
\bibitem{NoteCompCloud}Alternatively, one could think of compressing the radial direction $\sigma$ only, but this prompts the mean-field to increase $\eta$. The elongated condensate then acts like a waveguide for scattered photons and the impact of Mie scattering is reduced. The only way seems to be to reduce the interaction in the vicinity of a Feshbach resonance.
\bibitem{NoteMultiScat}Note that at $\Delta=(2\pi)~500~$MHz for high atom numbers, $N>10^6$, multiple scattering starts to set in and may weakly affect the data.
\end{thebibliography}
\end{document}